\documentclass[reqno]{amsart}
\usepackage[all]{xy}
\usepackage[dvips]{graphics,psfrag}

\def \Hil {\mathcal{H}}

\def \D {\mathcal{D}}

\def \R {\mathbb{R}}

\def \C {\mathbb{C}}
\def\u{\mathfrak{u}}
\def\gl{\mathfrak{gl}}

\def \PH {\mathcal{PH}}
\def \CP {\mathbb{CP}}

\def \Tr {\mathrm{Tr}}

\def \Hil {\mathcal{H}}

\def \D {\mathcal{D}}

\def \R {\mathbb{R}}

\def \C {\mathbb{C}}
\def\u{\mathfrak{u}}

\def \PH {\mathcal{PH}}
\def \RH {\mathcal{RH}}
\def \CP {\mathbb{CP}}

\def \Tr {\mathrm{Tr}}

\newcommand{\pd}[1]{\frac{\partial }{\partial #1}}

\newcommand{\scalar}[1]{\langle #1 \rangle }

\newtheorem{theorem}{Theorem}

\newtheorem{proposition}{Proposition}
\newtheorem{definition}{Definition}
\newtheorem{lemma}{Lemma}

\begin{document}

\title{The space of density states in geometrical quantum mechanics}
\author{Jes\'us Clemente-Gallardo}
\address{BIFI-Universidad de Zaragoza \\Corona de Arag\'on 42,  50009 Zaragoza-SPAIN}
\author{Giuseppe Marmo}
\address{Dipartamento di Scienze Fisiche \\ Universit\'a Federico II and INFN-
  Sezione Napoli \\ Via Cintia I-80126 Napoli-ITALY}

\begin{abstract}
We present a geometrical description of the space of density states of a
quantum system of finite dimension. After presenting a brief summary of the
geometrical formulation of Quantum Mechanics, we proceed to describe the space
of density states $\D(\Hil)$ from a geometrical perspective identifying the stratification
associated to the natural $GL(\Hil)$--action on $\D(\Hil)$ and some of its
properties. We apply this construction to the cases of quantum systems of two
and three levels.
\end{abstract}



\maketitle


\section{Introduction}
A comparison of the frameworks underlying classical and quantum mechanics shows
that the two descriptions  have several common mathematical structures. However,
a striking difference emerges: the classical setting is geometrical and
nonlinear while the quantum is algebraic and linear. The emphasis on the
underlying linearity in quantum mechanics is usually attributed to the
description of the interference phenomena \cite{Dirac:36}. Therefore, the carrier space of
quantum systems is required to be a Hilbert space $\Hil$ from the beginning. The
Hermitian structure is required to describe the probabilistic interpretation of
Quantum Mechanics. However, it is exactly this probabilistic interpretation
which forces  on us the identification of physical states not with the Hilbert
space but rather with the space of rays, i.e. the complex projective space of
$\Hil$, say $ \RH$. Of course $\RH$ is a genuine nonlinear manifold and on it
the Hermitian structure gives rise to a K\"ahler structure.

The appearance of this manifold in the quantum setting calls for a geometrical
formulation of Quantum Mechanics. It is clear that in the manifold view point
we  have to give up the usual ``superposition of states'' and the notion of
operators, eigenvectors and eigenstates as usually presented. Nevertheless, due
to their physical relevance and interpretation we must be able to recover these
``attributes'' for quantum systems also at the manifold level. The overall
formulation must allow for nonlinear transformations and therefore only
tensorial objects should be identified with physically relevant quantities. To
fully exploit the geometrical picture, one prefers to work with real
differential manifolds, i.e. one replaces the complex vector space $\Hil$ with
its realification $\Hil_\R$.  The Hermitian structure then splits into a
complex structure, a symplectic structure and a Riemannian structure
(compatible among them to define a K\"ahler structure) Hermitian operators are
transformed into functions by replacing them by their expectation values. These
functions project onto $\RH$. With the help of the Poisson tensor associated
with the symplectic structure it is possible to give rise to a flow by
integrating the Hamiltonian vector field associated with the expectation value
function corresponding to a given Hermitian operator
\cite{cirelli-3,anandan,ashschi:1999,
benvegnu,bloch,brody,cirelli2,cirelli3,darius,heslot,rowe}. 

The symplectic structure appearing in Quantum Mechanics makes also possible to
consider it as a ``classical field theory'' associated with a Lagrangian
description  with relativity group  the Galilei group \cite{marmovilasi},
since we deal with non-relativistic Quantum Mechanics. Thus, classical and
quantum descriptions have in common a symplectic structure. However, this is
the only quantum feature that has a direct classical analogue. Some characteristic
features like the quantum uncertainties and state vector reduction in a
measurement process are strictly related to the additional complex structure,
available in Quantum Mechanics but not present in Classical Mechanics
\cite{marmoscolarici}. This additional structure lies at the heart of the difference
between the 
mathematical structures underlying the two theories, much more than the linear
structure. 

Going back to the manifold view point introduced in Quantum Mechanics, i.e. the
identification of $\RH$ as the true manifold of physical states, we have to
recover the notion of superposition of physical states \cite{mankoz} . This has been done and 
creates a deep relation with Pancharatnam connection, Bargmann invariants and
geometric phases \cite{mukunda}. In addition we have to recover the notion of
``eigenvector'' and ``eigenvalue''. As a matter of fact by considering the
expectation value function associated with any operator we find that their
critical points will correspond to eigenstates and their values at those
critical points correspond to the eigenvalues. If we consider the expectation
value-functions of generic operators we get complex valued functions on $\RH$
and they provide us with a $\C^*$ --algebra, thus paving the way for the
geometrization of the $\C^*$--algebraic approach to quantum theories. In this
latter approach usually the space of states is identified with the space of
normalized positive linear functionals on the $\C^*$--algebra describing the
quantum system. As a provisional geometrization of this approach we shall
construct these spaces with the help of the momentum map associated with the
symplectic action of the unitary group on the K\"ahler manifold $\RH$. 

\section{A brief exposition of geometrical quantum mechanics}

The aim of this section is to present a brief summary of the set of the
geometrical tools which characterize the description of Quantum Mechanics
\cite{Grabowski:2005,Grabowski:2006,manko}. 

\subsection{The states}
The first step consists in replacing the usual complex vector space structure
of the Hilbert space $\Hil$ of a quantum system by the corresponding
realification of the  vector
space. We shall denote as $\Hil_\R$ the resulting vector space.
In this realification process the complex structure on
$\Hil$ will be represented by a tensor $J$ on $\Hil_\R$.

The natural identification is then  provided by
$$
\psi_R+i\psi_I=\psi \in \Hil \mapsto (\psi_R,\psi_I)\in \Hil_\R.
$$
Under this transformation, the Hermitian product becomes, for $\psi^1, \psi^2\in
\Hil$ 
$$
\langle(\psi^1_R,\psi^1_I), (\psi^2_R,\psi^2_I)\rangle =(\langle \psi^1_R, \psi^2_R\rangle + \langle \psi^1_I,
\psi^2_I\rangle )+i( \langle \psi^1_R, \psi^2_I\rangle -\langle \psi^1_I,\psi^2_R\rangle ).
$$

To consider $\Hil_\R$ just as a real differential manifold, the algebraic
structures available on $\Hil$ must be converted into tensor fields on
$\Hil_\R$. To this end we have to introduce the tangent bundle $T\Hil_\R$ and
its dual the cotangent bundle $T^*\Hil_\R$. The linear structure available in
$\Hil_\R$ is encoded in the vector field $\Delta$
$$
\Delta:\Hil_\R\to T\Hil_\R \quad \psi \mapsto (\psi, \psi)
$$

We can consider the Hermitian structure on $\Hil_R$ as an Hermitian
tensor on $T\Hil_\R$. With every vector we can associate a vector field
$$
X_\psi:\phi \to (\phi, \psi)
$$

Therefore, the Hermitian tensor, denoted in the same way as the scalar product
will be
$$
\langle X_{\psi_1}, X_{\psi_2}\rangle =\langle \psi_1, \psi_2\rangle 
$$
 The scalar product above is written as 
$
\langle \psi_1, \psi_2\rangle=g(X_{\psi_1}, X_{\psi_2})+i\,\omega (X_{\psi_1}, X_{\psi_2}),
$
where $g$ is now a symmetric tensor and $\omega $ a skew-symmetric one.  It is also
possible to write them as a pull-back by means of the dilation vector field $\Delta$ as:
$$
(\Delta^*(g+i\omega))(\psi, \phi)=\langle \psi, \phi \rangle_ {\Hil}
$$

The properties of the Hermitian product ensure that:
\begin{itemize}
\item the symmetric tensor is positive definite and non-degenerate, and hence
  defines a Riemannian structure on the real vector space.
\item the skew-symmetric tensor is also non degenerate, and is closed with
  respect to the natural differential structure of the vector space. Hence, the
  tensor is a symplectic form.
\end{itemize}

As the inner product is sesquilinear, it satisfies
$$
\langle \psi_1, i\psi_2\rangle =i\langle \psi_1, \psi_2\rangle, \qquad \langle i\psi_{1}, \psi_{2}\rangle =-i\langle
\psi_{1},\psi_{2}\rangle. 
$$
This  implies
$$
g(X_{\psi_1}, X_{\psi_2})=\omega (JX_{\psi_1}, X_{\psi_2}).
$$
We also have that $J^2=-\mathbb{I}$, and hence that the triple $(J, g, \omega )$
defines a K\"ahler structure.  This implies, among other things, that the tensor
$J$ generates both finite and infinitesimal transformations which are
orthogonal and symplectic.

Linear transformations are converted into $(1,1)$--tensor fields by setting $A\to
T_A$ where
$$
T_A:T\Hil_\R\to T\Hil_\R \quad (\psi, \phi)\mapsto (\psi, A\phi).
$$
The association $A\to T_A$ is an associative algebra isomorphism. It is possible
to recover the Lie algebra of vector fields by setting $X_A=T_A(\Delta)$.
Complex linear transformations will be represented by $(1,1)$--tensor fields
commuting with $J$.

For finite dimensional Hilbert spaces it may be convenient to
introduce adapted coordinates on $\Hil$ and $\Hil_\R$.
Fixing an orthonormal basis $\{ |e_k\rangle \} $ of the Hilbert space allows us to
identify this product with the canonical Hermitian product of $\C^n$:
$$
\langle \psi_1, \psi_2\rangle =\sum_k\langle \psi_1, e_k\rangle \langle e_k, \psi_2\rangle 
$$
The group of unitary transformations on $\Hil$ becomes identified with the
group $U(n, \C)$, its Lie algebra $\mathfrak{u}(\Hil)$ with $\mathfrak{u}(n,
\C)$ and so on.

The choice of the basis also allows us to introduce coordinates for the
realified structure:
$$
\langle e_k, \psi\rangle =(q_k+ip_k)(\psi),
$$ 
and write the geometrical objects introduced above as:
$$
J=\partial_{p_k}\otimes dq_k-\partial_{q_k}\otimes dp_k 
\quad
g=dq_k\otimes dq_k+dp_k\otimes dp_k
\quad
\omega=dq_k\land dp_k
$$

If we combine them in complex coordinates we can write the Hermitian structure
in a simple way $z_n=q_n+ip_n$:
$$
h=\sum_kd\bar z_k\otimes dz_k
$$

In an analogous way we can consider a contravariant version of these
tensors. It is also possible to build it by using the isomorphism $T\Hil_\R\leftrightarrow
T^*\Hil_\R$ associated to the Riemannian tensor $g$. The result in both cases
is a K\"ahler structure for the dual vector space $\Hil_\R^*$ with the dual
complex structure $J^*$, a Riemannian tensor $G$ and a (symplectic) Poisson
tensor $\Omega$: The coordinate expressions with respect to the natural base are: 
\begin{itemize}
\item the Riemannian structure  $
G=\sum_{k=1}^n\left(\pd{q_k}\otimes \pd{q_k}+\pd{p_k}\otimes \pd{p_k}\right),
$ 
\item  the Poisson tensor
$
\Omega=\sum_{k=1}^n\left(\pd{q_k}\land \pd{p_k}\right)
$
\item while the complex structure has the form 
$$J=\sum_{k=1}^n\left(\pd{p_k}\otimes d{q_k}-\pd{q_k}\otimes d{p_k}\right)$$
\end{itemize}

\subsection{The observables}
The space of observables (i.e. of self-adjoint operators acting on $\Hil$) may
be identified with the dual $\mathfrak{u}^*(\Hil)$ of the real Lie algebra
$\mathfrak{u}(\Hil)$, according to the pairing between the unitary Lie algebra
and its dual given by
$$
A(T)=\frac i2 \Tr AT
$$

Under the previous isomorphism, $\mathfrak{u}^*(\Hil)$ becomes a Lie algebra
with product  defined by
$$
i[A, B]=[A, B]_-=(AB-BA)
$$

We can also transfer the Jordan product:
$$
[A,B]_+=2A\circ B=AB+BA
$$
Both structures are compatible. As a result, $\mathfrak{u}^*(\Hil)$ becomes a
Jordan-Lie algebra (see \cite{Emch,Lands:1998}).

We can also define a suitable scalar product, given by:
$$
\langle A, B\rangle=\frac 12 \Tr AB
$$
which turns the space into a real Hilbert space. This scalar product is the
restriction of the one on $\mathfrak{gl}(\Hil)$ defined as $\langle M, N\rangle =\frac 12
\Tr M^\dagger N$.

Besides this scalar product is compatible with the Lie-Jordan structure in the
following sense:
$$
\langle [A,\xi], B\rangle _{\mathfrak{u}^*(\Hil)}=\langle A, [\xi, B]\rangle _{\mathfrak{u}^*(\Hil)}
\quad
\langle [A,\xi]_+, B\rangle _{\mathfrak{u}^*(\Hil)}=\langle A, [\xi, B]_+\rangle _{\mathfrak{u}^*(\Hil)}
$$

These algebraic structures may be given a tensorial translation in terms of the
association $A\mapsto T_A$. However we can also associate complex valued functions with
linear operators $A\in \mathfrak{gl}(\Hil)$ by means of the scalar product
$$
\mathfrak{gl}(\Hil)\ni A\mapsto f_A=\frac 12 \langle \psi, A\psi\rangle_{\Hil}. 
$$

In more intrinsic terms we may write
$$
f_A=\frac 12 (g(\Delta, X_A)+i\omega(\Delta, X_A)).
$$
Hermitian operators give rise thus to quadratic real valued functions.

The association of operators with quadratic functions allows also to recover
the algebraic structures on $\u(\Hil)$ and $\u^*(\Hil)$ by means of
approapriate $(0,2)$--tensors on $\Hil_\R$.  By using the contravariant form of
the Hermitian tensor $G+i\Omega$ given by: 
$$
G+i\Omega=4 \frac{\partial}{\partial{z_k}}\otimes \frac{\partial}{\partial{\bar
  z_k}}=\frac{\partial}{\partial{q_k}}\otimes \frac{\partial}{\partial{q_k}}+
\frac{\partial}{\partial{p_k}}\otimes \frac{\partial}{\partial{p_k}}+
i \frac{\partial}{\partial{q_k}}\land \frac{\partial}{\partial{p_k}},
$$

it is possible to define a bracket
$$
\{ f,h\} _{\Hil}=\{ f,h\}_{g}+i\{ f,h\} _{\omega}
$$

In particular, for quadratic real valued functions we have
$$
\{ f_A,f_B\}_g =f_ {AB+BA}=2f_{A\circ B} \quad
\{ f_A,f_B\}_{\omega}=-if_{AB-BA}
$$

The imaginary part, i.e. $\{ \cdot, \cdot\} _\omega$, defines a Poisson bracket on the space of
functions. Both brackets allow us to define a tensorial version of the
Lie-Jordan  algebra of the set of operators.

For Hermitian operators we recover previously defined vector fields:
$$
\mathrm{grad}f_A=\widetilde A; \quad \mathrm{Ham} f_A=\widetilde {iA}
$$
where the vector fields associated with operators, we recall,  are defined by:
$$
\widetilde A:\Hil_{\R}\to T\Hil_{R} \quad \psi\mapsto (\psi, A\psi)
$$

$$
\widetilde{iA}:\Hil_\R\to T\Hil_{R} \quad \psi\mapsto (\psi, JA\psi) 
$$

We can also consider the algebraic structure associated to the full bracket $\{
\cdot, \cdot \}_\Hil $, as we associated above the Jordan product and the commutator of
operators to the brackets $\{ \cdot, \cdot\} _g$ and $\{ \cdot, \cdot\} _\omega$ respectively. It is
simple to see that it corresponds to the associative product of the set of
operators, i.e.
$$
\{ f_A, f_B\} _\Hil=\{ f_A, f_B\} _g+i \{ f_A, f_B\} _\omega=f_{AB+BA}+if_{AB-BA}=
2f_{AB}
$$

This particular bilinear product on quadratic functions may be written also as
a star product
$$
\{ f_A, f_B\} _\Hil=2f_{AB}=\langle df_A, df_B\rangle_{\Hil^*}=f_A\star f_B
$$ 
The set of quadratic functions endowed with such a structure turns out to be a
$\C^*$--algebra.

We see then that we can reconstruct all the information of the algebra of
operators starting only with real-valued functions defined on $\Hil_\R$. We
have thus
\begin{proposition}
  The Hamiltonian vector field $X_f$ (defined as $X_f=\hat \Omega(df)$) is a Killing
  vector field for the Riemannian tensor $G$ if and only if $f$ is a quadratic
  function associated with an Hermitian operator $A$, i.e. there exists $A=A^\dagger
  $ such that $f=f_A$. 
\end{proposition}

Finally, we can consider the problem of how to recover the eigenvalues and
eigenvectors of the operators at the level of the functions of $\Hil_R$. It is
simple to see that
\begin{itemize}
\item eigenvectors correspond to the critical points of functions $f_A$, i.e.
$$
df_A(\psi_*)=0 \text{ iff  $\psi_*$ is an eigenvector of $A$ }
$$
\item the corresponding eigenvalue is recovered by the value 
$$
\frac{f_A(\psi_*)}{\scalar{\psi_*, \psi_*}}
$$
\end{itemize}

Thus we can conclude that the K\"ahler manifold $(\Hil_\R, J, \omega, g)$
contains all the information of the usual formulation of Quantum Mechanics on a
complex Hilbert space.

Up to now we have concentrated our attention on states and observables. If we
consider observables as generators of transformations, i.e. we consider the
Hamiltonian flow associated to the corresponding functions, the invariance of
the tensor $G$ implies that the evolution is actually unitary. It is,
therefore, natural, to consider the action of the unitary group on the
realification of the complex vector space.

\section{The momentum map: geometrical structures on 
$\mathfrak{g}^*$} 
The unitary action of $U(\Hil)$ on $\Hil$ induces a symplectic action on the
symplectic manifold $(\Hil_\R, \omega)$.  By using the association
$$
F:\Hil_\R\times \u(\Hil)\to \R \quad (\psi, A)\mapsto \frac 12 \langle \psi, A\psi\rangle =f_{iA}(\psi),
$$
we find, with $F_A=f_{iA}:\Hil_\R\to \R$, that
$$
\{ F(A), F(B)\}_\omega =iF([A,B]).
$$
Thus if we fix $\psi$, we have a mapping $F(\psi):\u(\Hil)\to \R$. Thus with any
element $\psi \in \Hil$ we have an element in $\u^*(\Hil)$. Hence it defines a
momentum map  
$$
\mu:\Hil\to \mathfrak{u}^*(\Hil),
$$
which provides us with a symplectic realization of the natural Poisson manifold
structure available in $\u^*(\Hil)$. 
We can write the momentum map from $\Hil_\R$ to $\mathfrak{u}^*(\Hil)$ as
$$
\mu(\psi)=|\psi\rangle \langle \psi|
$$

If we make the convention that the dual
$\u^*(\Hil)$ of the (real) Lie algebra $\u(\Hil)$ is identified with Hermitian
operators by means of a scalar product, the product pairing between Hermitian
operators $A\in \u^*(\Hil)$ and the anti-Hermitian element $T\in \u(\Hil)$ will be
given by
$$
A(T)=\frac i2 \Tr (AT)
$$

If we denote the linear function on $\u^*(\Hil)$ associated with the element
$iA\in \u(\Hil)$ by $\hat A$, we have
$$
\mu^*(\hat A)=f_A
$$

The pullback of linear functions on $u^*(\Hil)$ is given by the quadratic
functions on $\Hil_\R$ associated with the corresponding Hermitian operators.

It is possible to show that the contravariant tensor fields on $\Hil_\R$
associated with the Hermitian structure are $\mu$--related with a complex
tensor on $\u^*(\Hil)$:
$$
\mu_*(G+i\Omega)=R+i\Lambda; 
$$
where the two new tensors $R$ and $\Lambda$ are defined by
$$
R(\xi)(\hat A, \hat B)=\scalar{\xi, [A,B]_+}_{\u^*}=\frac 12 \Tr (\xi (AB+BA))
$$
and
$$
\Lambda(\xi)(\hat A, \hat B)=\scalar{\xi, [A,B]_-}_{\u^*}=\frac 1{2i} \Tr (\xi (AB-BA))
$$

Clearly,
$$
G(\mu^*\hat A, \mu^*\hat B)+i\Omega(\mu^*\hat A, \mu^*\hat B)=\mu^*(R(\hat A, \hat
B)+i\Lambda(\hat A, \hat B)).
$$

As we know that $\u^*(\Hil)$  is foliated by symplectic manifolds, we wish to
consider more closely the map from  $\Hil_\R$ to the minimal symplectic orbit
on $\u^*(\Hil)$.

\section{The complex projective space}

As we have already remarked, the association of states of the quantum system
with vectors in the Hilbert 
space needs further qualifications because of the probabilistic interpretation
required in Quantum Mechanics. More specifically, states should be identified
with rays in the Hilbert space, i.e. equivalence classes of vectors, orbits of
non-null vectors under the action of $\C_0=\C-\{ 0\} $. The equivalence class of
the vector $\psi\in \Hil$  will be denoted then as
$$
[\psi]=\{ \lambda \psi , \lambda\in \C_0\} 
$$

As the infinitesimal generators of the real and imaginary parts of the action
of $\C_0$ on $\Hil_\R$  are given by the dilation vector field $\Delta$ and the
vector field $J(\Delta)$ respectively, it is clear that we have to undertake the
projection of the relevant tensors on $\Hil_\R$ to the complex projective space
or ray space $\mathcal{RH}$. 

Without entering in too many details, we find that we have to modify $G$ and
$\Omega$ by a conformal factor to turn them into projectable tensors. Specifically
we have
$$
\tilde G=g(\Delta, \Delta)G, \quad \tilde \Lambda =g(\Delta, \Delta)\Lambda.
$$

These tensors are projectable onto non-degenerate contravariant tensors on
$\mathcal{RH}$ and gives rise to a Lie-Jordan algebra structure on the space of
real valued functions whose Hamiltonian vector fields are also Killing vector
fields for the projection $\tilde G$.

As a matter of fact a theorem by Wigner allows us to state that these functions
are necessarily projections of expectation values of Hermitian operators
$$
e_A(\psi)=\frac{\scalar{\psi, A\psi}}{\scalar{\psi,\psi}}.
$$

The action of the unitary group  may also be projected and gives rise
to a symplectic action on $\mathcal{RH}$. The momentum map from $\Hil_\R$
projects onto the momentum map from $\mathcal{RH}$ because it is equivariant
with respect to the action of $\Delta_\Hil$ on $\Hil_\R$ and the action of
$\Delta_{\u^*}$ on $\u^*(\Hil)$.

From $\mu^*(\hat A)=f_A$ we find
$$
\Delta_{\Hil} \mu^*(\hat A)=2\mu^*(\Delta_{\u^*}\hat A)=2\mu^*(\hat A) 
$$

The momentum map for the projected action may be written in the form
$$
\mu ([\psi])=\frac{|\psi \rangle \langle \psi | }{\scalar{\psi, \psi}}=\rho_\psi.
$$

This map identifies $\mathcal{RH }$ with the Hermitian operators in
$\u^*(\Hil)$  which are of rank one and are projectors, i.e.
$$
\rho_\psi \rho_\psi=\rho_\psi , \quad \Tr \rho_\psi=1.
$$

As the ray space is a principal bundle with base manifold $\mathcal{RH}$ and
structure group $\C_0$, we may look for a connection one form. 

The connection one-form $\theta$ is given by
$$
\theta(\psi)=\frac{\scalar{\psi, d\psi}}{\scalar{\psi, \psi}},
$$
with associated curvature form $\omega=d\theta$, because the structure group is
Abelian. This curvature form coincides with the symplectic structure on
$\mathcal{RH}$ arising from the projection of $\tilde \Lambda$ (conformally related
to $\Lambda$). 

It is also possible to write the Hermitian tensor which coincides with
the Hermitian tensor on $\mathcal{RH}$ when evaluated on horizontal vector
fields. We thus have
$$
\frac{\scalar{d\psi, d\psi}}{\scalar{\psi, \psi}}-\frac{\scalar{\psi, d\psi}\scalar{d\psi,
    \psi}}{\scalar{\psi, \psi}^2}.
$$

It is not difficult to see that both $\Delta$ and $J(\Delta)$ are annihilated by this
tensor. 

The embedding of $\mathcal{RH }$ into $\u^*(\Hil)$ by means of the momentum map
allows us to consider convex combinations of the image $\mu (\mathcal{RH})\subset
\u^*(\Hil)$. The convex combinations will generate the space of density states,
i.e. normalized positive linear functionals on the Lie-Jordan algebra of
observables.  This convex body inherits some structures from those existing on
$\u^*(\Hil)$, which are particularly important and  useful when we are interested in
describing evolutions of states which are not unitary.  In particular they
inherits a Poisson structure and a Jordan structure. In the next section  we
shall study more closely the space of density states.

\section{The space of density states}

As we have already remarked the space of density states is the space of
positive normalized  linear functionals on the real linear space of
observables. A theorem by Gleason \cite{gleason} asserts that these functionals may be
represented by suitable  operators when the trace is used as a bilinear
pairing. By using this theorem we can start by considering  states directly as
appropriate operators.  

Let us introduce first the space of all non-negatively defined operators,
i.e. the space of all those $\rho\in \mathfrak{gl}(\Hil)$  which can be written
in the form
$$
\rho=T^\dagger T\quad T\in \gl(\Hil).
$$
We shall denote by $\mathcal{PH}$ this space of operators, which is a convex
cone in $\u^*(\Hil)$. By imposing the condition $\Tr \rho=1$  we select in $\PH$
the convex body of density states which we denote by $\D(\Hil)$. We shall also
consider non-negative Hermitian operators and density states of rank $k$, and
denote as $\mathcal{P}^k(\Hil)$ and $\D^k(\Hil)$ respectively the corresponding
spaces. 

The complex projective space is in one-to-one correspondence with
$\D^1(\Hil)$. Indeed, any state in $\D(\Hil)$ can be written as a convex
combination of distinct states $\rho=p_1\rho_1+(1-p_1)\rho_2$, with $0\leq p_1\leq 1$.
We shall call {\bf extremal states} those which can not be written in this
form (i.e. as convex combination of two $\rho_1$ and $\rho_2$). The extremal
states are thus given by $\D^1(\Hil)$.

As $\Lambda$ and $R$ are not invertible in $\u^*(\Hil)$, it is convenient to use the
pairing between $\u(\Hil)$ and $\u^*(\Hil)$ defined by the trace, to introduce
two tensor fields on $\u^*(\Hil)$. We set then

$$
\tilde J,R:T\u^*(\Hil)\to T\u^*(\Hil)
$$
defined as
\begin{align*}
\tilde J_\xi(X_A)=(\xi, [A, \xi]_-)=\Lambda (\xi)(d\hat A) \\
R_\xi(X_A)=(\xi, [A, \xi]_+)=R (\xi)(d\hat A) 
\end{align*}

The image of $\tilde J$ is the Hamiltonian  involutive distribution associated with
linear Hamiltonian functions, and we shall denote it as $D_\Lambda$. The image of $R$
is also a distribution, which we shall denote as $D_R$, but in this case it is
not involutive.  It is possible to see that combining $D_R$ and $D_\Lambda$ we
can define two distributions $D_0=D_R\cap D_A$ and $D_1=D_R+D_\Lambda$ which are indeed
involutive. 

We notice that the tensors $\tilde J$ and $R$ commute, i.e. $\tilde J\circ R=R\circ
\tilde J$. More specifically we have
$$
\tilde J(\xi)\circ R(\xi)(X_A)=R(\xi )\circ \tilde J(\xi)(X_A)=[A, \xi ^2]_-.
$$
As a result, we find that the distribution $D_0$ becomes:
$$
D_0(\xi)=\{ [A, \xi^2]; A\in \u^*(\Hil)\}. 
$$
On $\mathcal{RH}$, $D_0$ coincides with $D_\Lambda$.

The distribution $D_1$ is involutive and the leaves are related to orbits of
the following $GL(\Hil)$--action:
$$
GL(\Hil)\times \u^*(\Hil)\to \u^*(\Hil)\quad (T, \xi)\mapsto T\xi T ^\dagger. 
$$

We obtain some interesting results \cite{Grabowski:2005,Grabowski:2006}:
\begin{enumerate}
\item The Hermitian  operators $\rho$ and $\rho '$ belong to the same $GL$--orbit if
  and only if they have the same number $K_+$ of positive eigenvalues and the
  same number $K_-$ of negative eigenvalues (counted with multiplicities).
\item Any $GL$--orbit intersecting the positive cone $\PH$ lies entirely in
  $\PH$; so that $\PH$ is stratified by the $GL$--orbits. These $GL$--orbits
  in $\PH$ are determined by the rank of the operator, i.e. they are exactly
  $\mathcal{P}^k(\Hil)$. 
\end{enumerate}

When we restrict to the space of density states by imposing the condition $\Tr
\rho=1$, this $GL$--action will not preserve the states. It is however possible to
define a new action that maps $\D(\Hil)$ into itself by setting
$$
GL(\Hil)\times \D (\Hil)\to \D(\Hil) \quad (T, \rho)\mapsto \frac{T\rho T^\dagger}{\Tr (T\rho T^\dagger)}.
$$

This action does preserve the rank of $\rho$ and then the following proposition
holds true:
\begin{proposition}
  The decomposition of the convex body of density states $\D(\Hil)$ into orbits
  of the $GL(\Hil)$--action $\rho\mapsto \frac{T\rho T^\dagger}{\Tr (T\rho T^\dagger)}$ is exactly the
  stratification 
$$
\D(\Hil)=\bigcup_{k=1}^n \D^k(\Hil),
$$
into states of a given rank.
\end{proposition}

The boundary of the convex body of density states  consists of states of rank
lower than $n$, i.e. $\partial \D(\Hil)=\bigcup_{k=1}^{n-1} \D^k(\Hil)$, and each stratum is
a smooth submanifold in $\u^*(\Hil)$. However, the boundary $\partial \D(\Hil)$  is
not smooth (for $n> 2$). We have the following theorem:

\begin{theorem}
  Every smooth curve $\gamma:\R\to \u^*(\Hil)$ through the convex body of density
  states is tangent, at every point, to the stratum to which it belongs, i.e.
$$
\gamma(t)\in \D^k(\Hil) \Rightarrow T\gamma(t)\in T_{\gamma(t)}\D^k(\Hil).
$$
\end{theorem}

One may gain further insight on the ``location'' of the boundary by using the
notion of ``face''
\begin{definition}
  A non-empty closed convex subset $K_0$  of a closed convex set $K$ is called
  a {\bf face} of $K$ if any closed segment in $K$ with an interior point in
  $K_0$ lies entirely in $K_0$. 
\end{definition}

Thus, for any $\rho\in \D(\Hil)$ we may consider the decomposition
$\Hil=\mathrm{Im}\rho +\mathrm{Ker}\rho$ into the kernel and the image of $\rho$. We
have: 
\begin{proposition}
  The face of $\D(\Hil)$ through $\rho\in \D^k(\Hil)$ consists of states $A$
  which are ``projectable'' with respect to the projection defined by
  $\mathrm{Ker}\rho$, i.e. $\mathrm{Ker} A\subset \mathrm{Ker}\rho$. The face through $\rho$
  is then equivalent to $\D(\mathrm{Im} \rho)$. 
\end{proposition}

The inner product defined by the trace allows to define a probability
transition function
$$
p(\rho_1, \rho_2)=\Tr \rho_1\rho_2,
$$
when $\rho_1$ and $\rho_2$ belong to the boundary $\partial_e \D(\Hil)$, the space of
extremal states.

This function satisfies
$$
0\leq p(\rho_1,\rho_2)\leq 1 \quad p(\rho_1, \rho_2)=p(\rho_2, \rho_1).
$$
Moreover, $p(\rho_1, \rho_2)=1$ if and only if $\rho_1=\rho_2$.

It is not difficult to show that the Hamiltonian vector fields which leave $R$
invariant will preserve also the probability transition functions. This result
is related to a theorem by Wigner and may be used to recover the space of
density states starting with a Poisson space carrying a compatible probability
transition function. Further details and a full treatement of Poisson spaces
with a transition probability function have been considered by Landsman (see
\cite{Lands:Poisson}). 

\section{Two examples: $\mathfrak{g}_{\mathfrak{u}(2)}$ and
  $\mathfrak{g}_{\mathfrak{u}(3)}$} 

\subsection{States of a two level system}
We shall consider in some detail two examples. The first one is the two level
system with carrier space $\Hil=\C^2$. We consider $\mathfrak{u}(2)$ and
$\mathfrak{u}^*(2)$ and make a specific choice of basis
$$
\sigma_0=
\begin{pmatrix}
1 & 0 \\
0 & 1
\end{pmatrix}
\quad 
\sigma_1=
\begin{pmatrix}
0 & i \\
-i & 0
\end{pmatrix}
\quad 
\sigma_2=
\begin{pmatrix}
0 & 1 \\
1 & 0
\end{pmatrix}
\quad 
\sigma_3=
\begin{pmatrix}
1 & 0 \\
0 & -1
\end{pmatrix} 
$$

We recall that
$$
\sigma_1\sigma_2=i\sigma_3 \quad \sigma_2\sigma_3=i\sigma_1 \quad \sigma_3\sigma_1=i\sigma_2,
$$
along with 
$$
\sigma_2\sigma_1=-i\sigma_3 \quad \sigma_3\sigma_2=-i\sigma_1 \quad \sigma_1\sigma_3=-i\sigma_2;
$$
which may be obtained considering the conjugate-transpose of any product.

We define coordinate functions by writing
$$
y_0(A)=\frac 12 \Tr (\sigma_0A), \quad y_a(A)=\frac 12 \Tr (\sigma_a A).
$$

In these coordinates the corresponding Poisson brackets for the canonical
Lie-Poisson structure on the dual of the Lie algebra read:
$$
\{ y_0, y_a\}=0 \quad \{ y_a, y_b\}=2\epsilon_{abc}y_c.
$$
The expression of the Poisson tensor thus becomes:
$$
\Lambda=2\left ( y_1\pd{y_2}\land \pd{y_3}+y_2\pd{y_3}\land \pd{y_1}+y_3\pd{y_1}\land
  \pd{y_2}\right )
$$

It is also possible to construct the Riemann-Jordan tensor in the form:
\begin{multline*}
R=\pd{y_0}\otimes_s\left (y_1\pd{y_1}+y_2\pd{y_2}+y_3\pd{y_3} \right ) +\\
y_0\left ( \pd{y_0}\otimes \pd{y_0}+\pd{y_1}\otimes \pd{y_1}+\pd{y_2}\otimes \pd{y_2}+\pd{y_3}\otimes
  \pd{y_3} \right )
\end{multline*}
where $\otimes_s$ means the symmetrized tensor product.

\subsection{Distributions associated with $\Lambda$ and $R$}

It is easy to see that the Hamiltonian distribution is generated by
$$
H_1=y_3\pd{y_2}-y_2\pd{y_3}, \quad H_2=y_1\pd{y_3}-y_3\pd{y_1}, \quad
H_3=y_2\pd{y_1}-y_1\pd{y_2},
$$
while the distribution associated with the Riemann-Jordan tensor is
$$
X_0=y^a\pd{y^a}+y^0\pd{y^0} \quad 
X_a=y^a\pd{y^0}+y^0\pd{y^a}
$$

It is clear that $X_0$ is central and $\{ X_a\} $ are boosts of a four
dimensional Lorentz group, therefore their commutator will provide us with the
Lie algebra of the rotation group:
$$
[X_a, X_b]=y^a\pd{y^b}-y^b\pd{y^a}.
$$

The intersection of the distribution associated with $R$ and the Hamiltonian
distribution associated will indeed be generated by the Hamiltonian vector
fields and is involutive with leaves which are symplectic two dimensional
spheres. The distribution  generated by the union of the two distributions is
the full Lorentz group centrally extended with the dilations. As the Lorentz
group admits as a covering $SL(2, \C)$ the central extension is isomorphic to
$GL(2, \C)$. This is a general property holding true in any dimension (see
\cite{Grabowski:2006}).

We find that 

\begin{lemma}
The rank of $\Lambda$ is zero if $y_1^2+y_2^2+y_3^2=0$ and the rank is
equal to 2 if  $y_1^2+y_2^2+y_3^2>0$.    
\end{lemma}

The situation is richer with $R$: 
\begin{lemma}
The rank
of $R$ is
\begin{itemize}
\item zero if $y_0^2+y_1^2+y_2^2+y_3^2=0$   
\item two if $y_0=0$ and $y_1^2+y_2^2+y_3^2>0$.
\item three for $y_0^2=y_1^2+y_2^2+y_3^2$
\item  four if $y_0^2\neq y_1^2+y_2^2+y_3^2$
\end{itemize}
 \end{lemma}

\subsection{Density states}

As we  have already seen in the previous sections the set of states is
identified with a subset of $\u^*(\Hil)$ satisfying a positivity condition and
a normalization condition.  In the specific situation we are considering, a
generic Hermitian matrix $A=y^0\sigma_0+y^a\sigma_a$ will define a state if
$$
\Tr A=1 \quad 0\leq y^0+y^3\leq 1, \quad 0\leq y^0-y^3\leq 1, \quad \det A\geq 0.
$$

Explicitly we have 
$$
y^0=\frac 12 , \quad \left (\frac 12 +y^3\right ) \left ( \frac 12 -y^3\right
)- ((y^1)^2+(y^2)^2)\geq 0,
$$
or
$$
(y^3)^2+(y^2 )^2+(y^1)^2\leq \frac 14.
$$

Thus in our parametrization states are determined by points in $\R^4$  on the
hyperplane $y^0=\frac 12 $, and on this three dimensional space are identified by
the points in the ball of radius $\frac 12$. When referring to states we
replace $A$ with $\rho$ and write:
\begin{equation}
  \label{eq-rho}
  \rho=
\begin{pmatrix}
\frac 12 +y^3 & y^2+iy^1 \\
y^2-iy^1 & \frac 12 -y^3
\end{pmatrix}.
\end{equation}

The pure states corresponding to the vector $(z_1, z_2)\in \C^2$ with unit norm
$z_1\bar z_1+z_2\bar z_2=1$ has a density state
$$
\rho=
\begin{pmatrix}
\bar z_1 \\
\bar z_2
\end{pmatrix}\otimes
(z_1, z_2)=
\begin{pmatrix}
z_1\bar z_1 & \bar z_1 z_2 \\
\bar z_2 z_1 & z_2\bar z_2
\end{pmatrix}.
$$

Within the previous parametrization we find 
$$
y^3=\frac 12 (z_1\bar z_1-z_2\bar z_2), \quad y^1=\mathrm{Im}(\bar z_1 z_2),
\quad
y^2=\mathrm{Re}(\bar z_1z_2),
$$
and for these points the inequality is saturated thus implying that they lie on
the surface of the ball of radius $\frac 12$. We shall denote the set of
density states  by $\D$. This set is the convex hull of the sphere of pure
states. For any $\rho\in \D$  there exist pure states $\rho_1$ and $\rho_2$ and
a positive number $p$ such that $\rho=p\rho_1+(1-p)\rho_2$.

These states, points on the surface, are in one-to-one correspondence with the
unit rays in $\C^2$ and the map is given by the momentum map associated with the
symplectic action of $U(2)$ on $\mathcal{RH}\sim \CP^1$. The ball of the density
states is foliated by symplectic leaves associated with the coadjoint action of
$U(2)$, which coincide also with the orbits of the $SU(2)$ group.

The analysis of these orbits may also be done by considering the orbits passing
through diagonal matrices, in other terms
$$
\rho=S \begin{pmatrix} a & 0 \\ 0 & b\end{pmatrix}S^\dagger \quad a+b=1 \quad a\geq 0,
\quad b\geq 0.
$$

We visualize the situation with the help of the following diagram: the segment
connecting $(\frac 12, \frac 12)$ with $(1,0)$ parametrizes the family of two
dimensional spheres. The point $(\frac 12, \frac 12)$ coincides with the center
of the ball and $(1,0)$ belongs to the outmost sphere of pure states.

\begin{center}
\resizebox{!}{6cm}{
\psfrag{a}{$a$}
\psfrag{b}{$b$}
\includegraphics{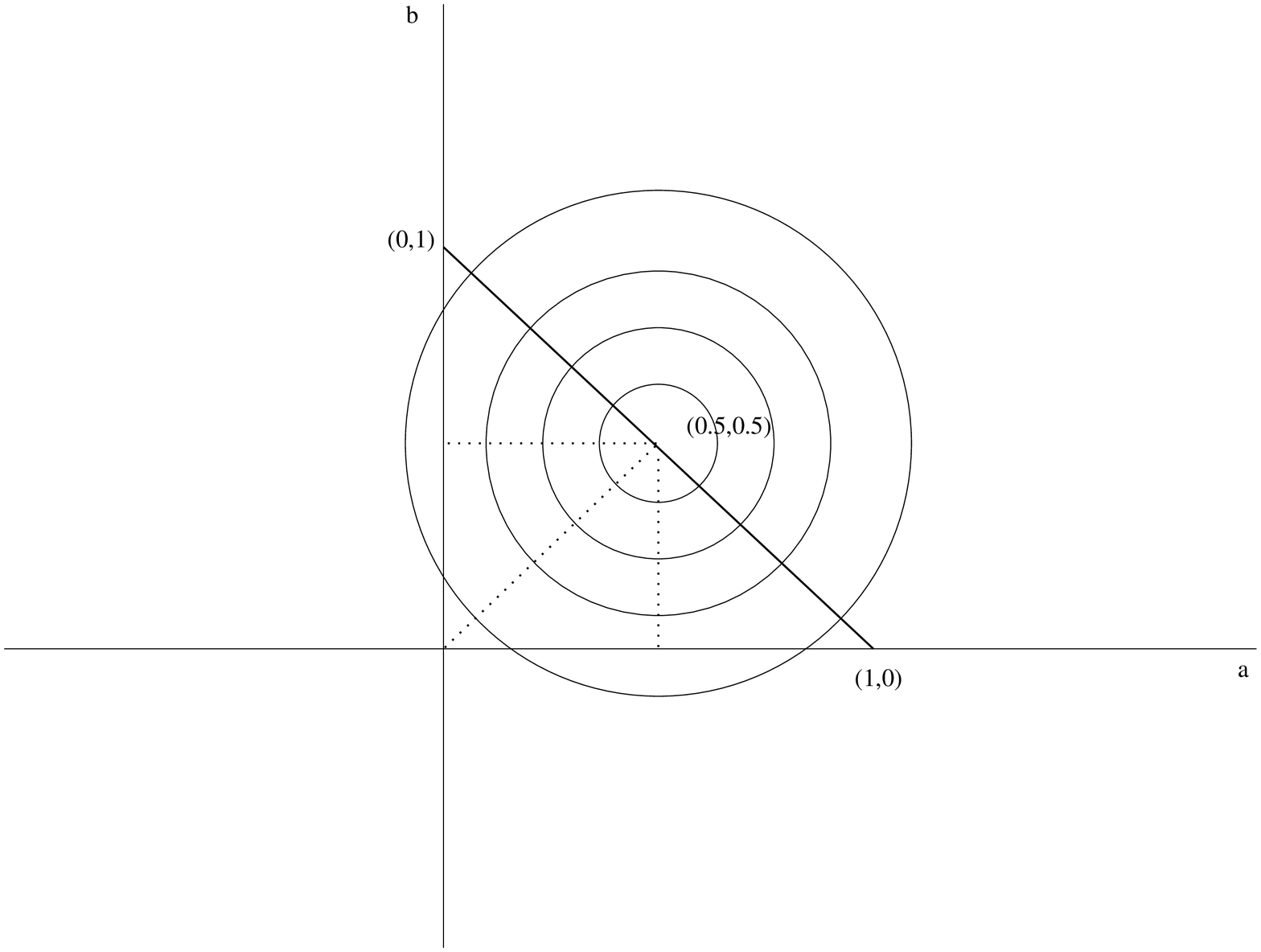}
}
\end{center}

What we have described is usually known as the Bloch sphere representation of
one qubit. The decomposition of a density states $\rho$, a point in the ball,
as a convex sum of two pure states $\rho_1=\frac{|\psi_1\rangle \langle \psi_1
  |}{\scalar{\psi_1,\psi_1}}$ and  $\rho_2=\frac{|\psi_2\rangle \langle \psi_2  |}{\scalar{\psi_2,\psi_2}}$, 
is given geometrically  by drawing a straight
  line through $\rho$: the states $\rho_1$ and $\rho_2$ are the intersections of
  the line with the sphere. Evidently this decomposition may be done in a two
  parameter family of ways.

\begin{center}
\resizebox{!}{6cm}{
\psfrag{y1}{$y_1$}
\psfrag{y2}{$y_2$}
\psfrag{y3}{$y_3$}
\psfrag{rho}{$\rho$}
\psfrag{rho1}{$\rho_1$}
\psfrag{rho2}{$\rho_2$}
\psfrag{rhop1}{$\rho'_1$}
\psfrag{rhop2}{$\rho'_2$}
\includegraphics{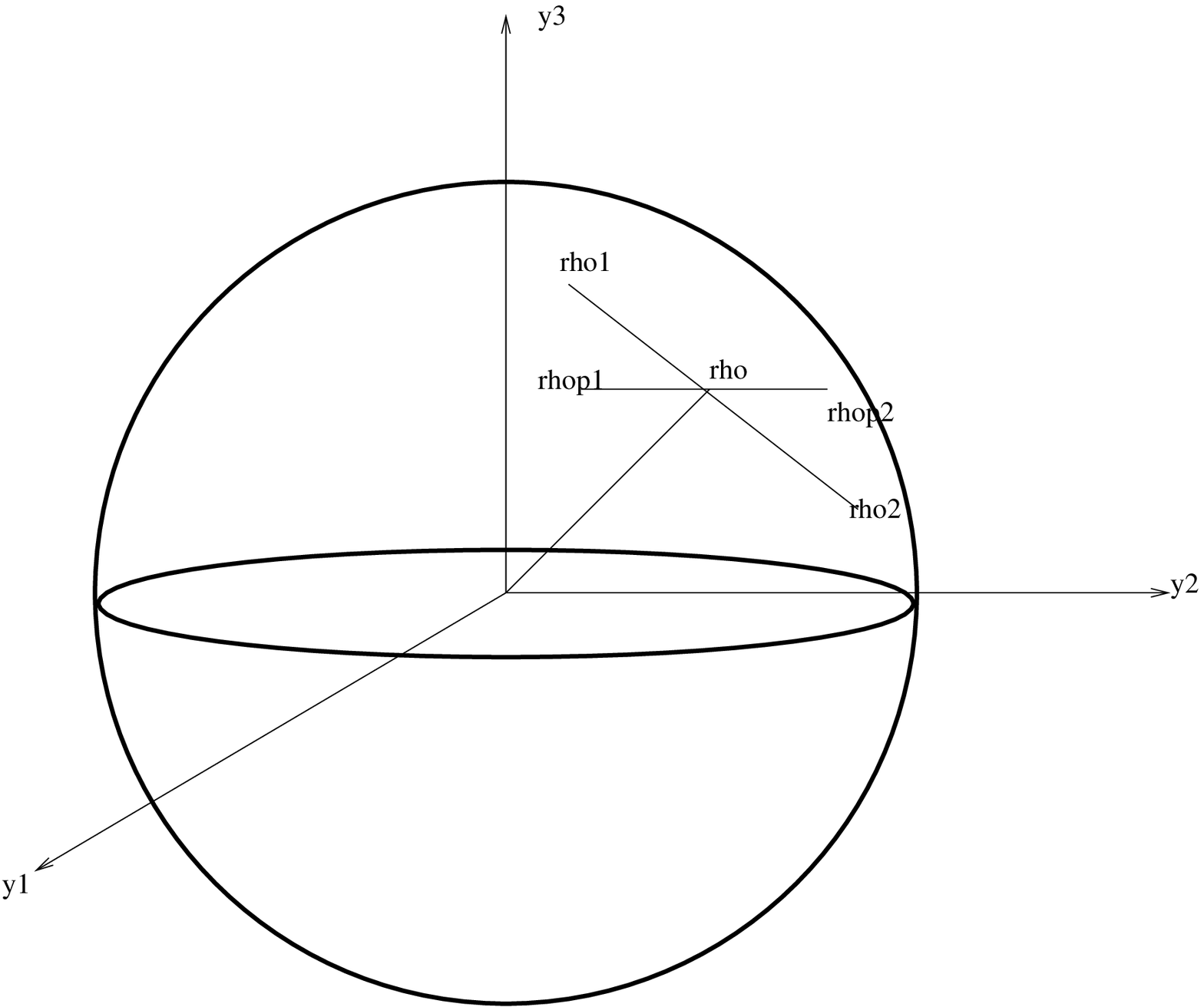}
}
\end{center}

\subsection{States of a three level system}

Now $\Hil=\C^3$. The states are normalized positive $3\times 3$ matrices inside
$\u^*(3)$. We first consider the geometrical tensors defined by means of the
momentum map construction. We choose a basis for $\u(3)$ given by the Gell-Mann
matrices 

$$
\lambda_1=
\begin{pmatrix}
0 & 1 & 0 \\
1 & 0 & 0 \\
0 & 0 & 0
\end{pmatrix}
\quad 
\lambda_2=
\begin{pmatrix}
0 & -i & 0 \\
i & 0 & 0 \\
0 & 0 & 0
\end{pmatrix}
\quad 
\lambda_3=
\begin{pmatrix}
1 & 0 & 0 \\
0 & -1 & 0 \\
0 & 0 & 0
\end{pmatrix}
$$
$$ 
\lambda_4=
\begin{pmatrix}
0 & 0 & 1 \\
0 & 0 & 0 \\
1 & 0 & 0
\end{pmatrix}
\quad
\lambda_5=
\begin{pmatrix}
0 & 0 & -i \\
0 & 0 & 0 \\
i & 0 & 0
\end{pmatrix}
\quad 
\lambda_6=
\begin{pmatrix}
0 & 0 & 0 \\
0 & 0 & 1 \\
0 & 1 & 0
\end{pmatrix}
$$
$$ 
\lambda_7=
\begin{pmatrix}
0 & 0 & 0 \\
0 & 0 & -i \\
0 & i & 0
\end{pmatrix}
\quad 
\lambda_8=\frac 1{\sqrt{3}}
\begin{pmatrix}
1 & 0 & 0 \\
0 & 1 & 0 \\
0 & 0 & -2
\end{pmatrix}
\quad 
\lambda_0=\sqrt{\frac 2 3}
\begin{pmatrix}
1 & 0 & 0 \\
0 & 1 & 0 \\
0 & 0 & 1
\end{pmatrix}
$$

They satisfy the scalar product relation
$$
\Tr \lambda_\mu \lambda_\nu=2\delta_{\mu \nu}
$$

Their commutation and anti-commutation relations are written in terms of the
antisymmetric structure constants and symmetric d--symbols $d_{\mu \nu \rho}$. We find
$$
[\lambda_\mu, \lambda_\nu]=2iC_{\mu \nu \rho}\lambda_\rho \quad
[\lambda_\mu, \lambda_\rho]_+=2\sqrt{\frac 2 3} \lambda_0\delta_{\mu \nu}+2d_{\mu \nu
  \rho}\lambda_\rho. 
$$

The numerical values turn out to be
\begin{multline*}
C_{123}=1, \qquad C_{458}=C_{678}=\frac{\sqrt{3}}2 \\
C_{147}=-C_{156}=C_{246}=C_{257}=C_{345}=-C_{367}=\frac 12
\end{multline*}
The values of these symbols show the different embeddings of $SU(2)$ into
$SU(3)\subset U(3)$.  For the other coefficients we have
\begin{multline*}
d_{jj0}=-d_{0jj}=-d_{j0j}=\sqrt{\frac 23} \quad j=1, \cdots , 8\\
-d_{888}=d_{8jj}=d_{jj8}=d_{j8j}=\frac 1{\sqrt{3}}\quad j=1, 2, 3 \\
d_{8jj}=d_{jj8}=d_{j8j}=-\frac 1{2\sqrt{3}}  \quad j=4,5,6,7\\
d_{3jj}=d_{jj3}=d_{j3j}=\frac 1{2}  \quad j=4,5 \qquad
d_{3jj}=d_{jj3}=d_{j3j}=-\frac 1{2}  \quad j=6,7\\
d_{146}=d_{157}=d_{164}=d_{175}=-d_{247}=d_{256}=d_{265}=-d_{274}=\frac 12 \\
d_{416}=-d_{427}=d_{461}=-d_{472}=d_{517}=d_{526}=d_{562}=d_{571}=\frac 12 \\
d_{614}=d_{625}=d_{641}=d_{652}=d_{715}=-d_{724}=d_{751}=-d_{742}=\frac 12
\end{multline*}

The indices appearing in the non-null structure constants are
identifying the corresponding $\lambda$--matrices whose pairwise commutators define
$SU(2)$--subgroups. It is now possible to introduce coordinate functions
$$
y^\mu(A)=\frac 12 \Tr \lambda_\mu A.
$$
In these coordinates, a generic Hermitian matrix $A$ can be written  as
$$
A=y^0\lambda_0+y^r\lambda_r
$$ 

The vector $(y^0, \vec y)\in \R^9$ plays a similar role to the one we saw on
$U(2)$. Under conjugation with $S\in SU(3)$, any matrix $A$ can be written as
$$
A=S
\begin{pmatrix}
a & 0 & 0 \\
0& b & 0 \\
0 & 0 & c
\end{pmatrix}
S^\dagger .
$$

The scalar product induced on vectors on $\R^8$ will be invariant under the
action of $SO(8)$. It is now possible to write the Poisson tensor 
$$
\Lambda=2C_{\mu \nu  \rho}y^\rho \pd{y^\mu}\land \pd{y^\nu}
$$
and the Riemann-Jordan tensor
$$
R=\pd{y^0}\otimes_sy^\mu \pd{y^\mu}+y^0\pd{y^r}\otimes \pd{y^r}+d_{\mu \nu
  \rho}y^\mu  \pd{y^\nu}\otimes_s\pd{y^\rho}.
$$

Now the analysis of the various distributions is more cumbersome, however it is
easy to identify a few elements:
$$
R(dy^0)=y^\mu \pd{y^\mu},
$$
which is the dilation vector field on $\R^9$; while
$R(dy^r)=y^r\pd{y^0}+y^0\pd{y^r}+d_{\mu  \nu r}y^\mu \pd{y^\nu}$, where it is possible
to identify a boost structure plus a correction due to the d--symbols. In any
case the union of the Hamiltonian distribution and the Riemannian-Jordan
distribution generates $GL(3, \C)$.

The set of states will again be identified as the subset of the Hermitian
matrices which are  normalized and satisfy the positivity condition. If we set
$$
\rho=
\begin{pmatrix}
a & \bar h & g \\
h & b & \bar f \\
\bar g & f & c
\end{pmatrix}
\quad a,b,c\in \R \quad f,g,h\in \C.
$$

The conditions for $\rho$ to be a state are:
\begin{itemize}
\item $a+b+c=1$
\item $a\geq 0$, $b\geq 0$, $c\geq 0$.
\item $|f|^2\leq bc $, $| g |^2\leq ca $, $|h|^2\leq ab$.
\item $\mathrm{det} \rho=abc+2\mathrm{Re}(fgh)-(a| f|^2+b|g|^2+c|h|^2)\geq 0$
\end{itemize}

These matrices form a convex set of $\R^8$. The trace condition allows to
identify this subset  as a subset of the vector space of $\R^8$ corresponding
to the dual space of the Lie algebra of $SU(3)$.

Extremal states are in one-to-one correspondence with the minimal symplectic
orbit of the unitary group according to the coadjoint action and corresponds to
$\CP^2$, the complex projective space of $\C^3$.

Pure states, rank one projectors, are given by vectors $(z_1, z_2,
z_3)\in \C^3$ with the normalization condition $z_1\bar z_1+z_2\bar z_2+z_3\bar
z_3=1$ as
$$
\begin{pmatrix}
z_1\bar z_1 & \bar z_1 z_2 & \bar z_1z_3\\
\bar z_2 z_1 & z_2\bar z_2 & \bar z_2 z_3\\
\bar z_3 z_1 & z_3\bar z_2 & \bar z_3 z_3
\end{pmatrix}
$$
Previous inequalities are saturated by these states.

These extremal states may be written in terms of the $\lambda$--matrices 
$$
\rho_\psi=\frac{| \psi \rangle \langle \psi| }{\scalar{\psi, \psi}}=\frac 13 (\mathrm{I}+\sqrt{3}n^a\lambda_a),
$$
with $n^an_a=1$ and  $n\star n=n$, the star product being 
$$
(a\star b)_l=\sqrt{3}d_{ljk}a_jb_k
$$

By using the ``radial-angular'' parametrization of states, say 
$$
\rho=s
\begin{pmatrix}
a & 0 & 0 \\
0 & b & 0 \\
0 & 0 & c
\end{pmatrix}
s^\dagger\quad
a\geq 0, b\geq 0, c\geq 0, \quad s\in SU(3), 
$$
we may study the structure of this union of symplectic orbits by considering
the family of diagonal matrices with the positivity condition (elements of a
positive Weyl chamber in the Abelian Cartan subalgebra). The hyperplane $\Tr
\rho=1$ identifies a triangle with the intersection with positive axes ($Oa, Ob,
Oc$); i.e. in the positive octant. 

\begin{center}
\resizebox{!}{8cm}{
\psfrag{a}{$a$}
\psfrag{b}{$b$}
\psfrag{c}{$c$}
\psfrag{0}{$0$}
\includegraphics{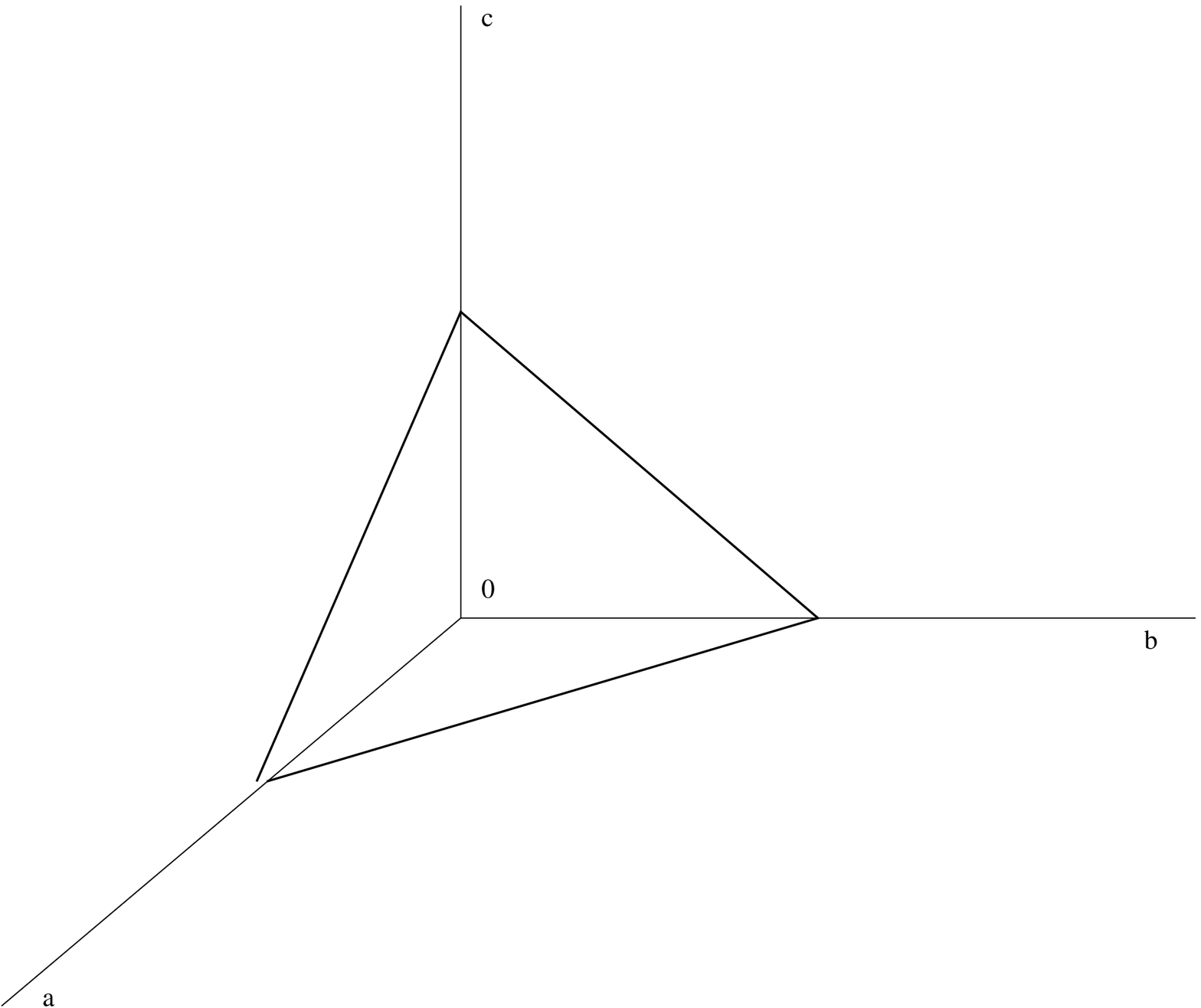}
}
\end{center}
Each  internal point of the triangle
corresponds to a 6--dimensional symplectic orbit, out of which we may consider
convex combinations. Due to the action of $SU(3)$ containing the action of the
discrete Weyl group, the symplectic orbits are actually parametrized by the
following smaller triangle.
\begin{center}
\resizebox{!}{8cm}{
\psfrag{a}{$a$}
\psfrag{b}{$b$}
\psfrag{c}{$c$}
\psfrag{0}{$0$}
\includegraphics{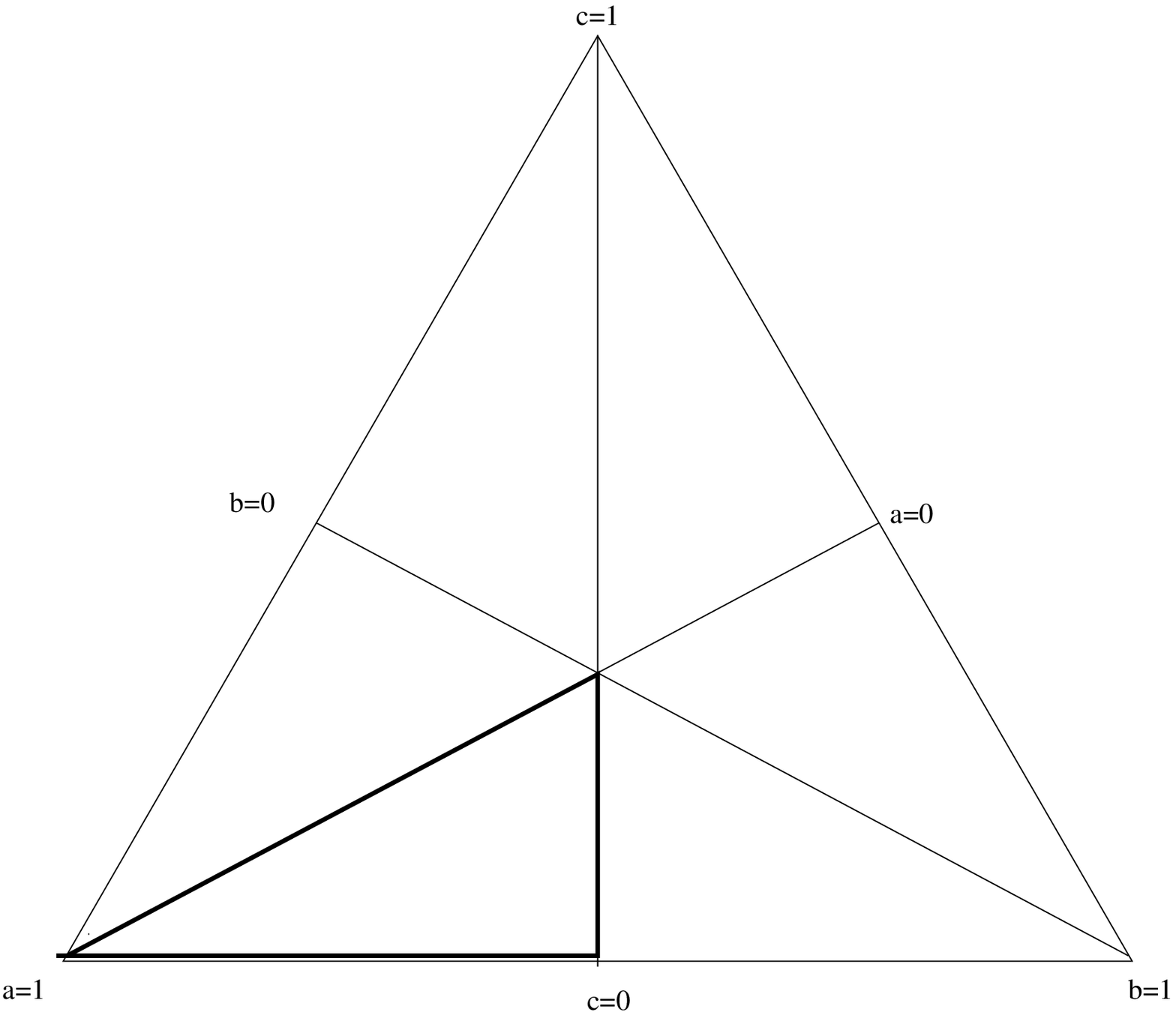}
}
\end{center}

When $a=b=c=\frac 13$ we have the ``maximally mixed state'' which play a
crucial role when we consider composite systems and entangled states (the orbit
passing through this point degenerates to a zero dimensional orbit). On the
boundary of the bigger triangle  the rank of $\rho$ is either 1 or 2. However the
orbits passing through these points are diffeomorphic to $\CP^2$. For a generic
point, the orbits are diffeomorphic to $SU(3)/U(1)\times U(1)$. It appears quite clearly
that the set of states is a stratified manifold characterized by the rank of
the state. We shall not indulge further on the geometrical analysis and refer
to the literature for further details and applications.

\end{document}